\documentstyle[11pt]{article}

\catcode`\@=11
\long\def\@makefntext#1{
\protect\noindent \hbox to 3.2pt {\hskip-.9pt
$^{{\ninerm\@thefnmark}}$\hfil}#1\hfill}		

\def\@makefnmark{\hbox to 0pt{$^{\@thefnmark}$\hss}}  
	
\def\ps@myheadings{\let\@mkboth\@gobbletwo
\def\@oddhead{\hbox{}
\rightmark\hfil\ninerm\thepage}
\def\@oddfoot{}\def\@evenhead{\ninerm\thepage\hfil
\leftmark\hbox{}}\def\@evenfoot{}
\def\sectionmark##1{}\def\subsectionmark##1{}}

\newcounter{sectionc}\newcounter{subsectionc}\newcounter{subsubsectionc}
\renewcommand{\section}[1] {\vspace*{0.6cm}\addtocounter{sectionc}{1}
\setcounter{subsectionc}{0}\setcounter{subsubsectionc}{0}\noindent
	{\normalsize\bf\thesectionc. #1}\par\vspace*{0.4cm}}
\renewcommand{\subsection}[1] {\vspace*{0.6cm}\addtocounter{subsectionc}{1}
	\setcounter{subsubsectionc}{0}\noindent
	{\normalsize\it\thesectionc.\thesubsectionc. #1}\par\vspace*{0.4cm}}
\renewcommand{\subsubsection}[1]
{\vspace*{0.6cm}\addtocounter{subsubsectionc}{1}
	\noindent  
{\normalsize\rm\thesectionc.\thesubsectionc.\thesubsubsectionc.
	#1}\par\vspace*{0.4cm}}
\newcommand{\nonumsection}[1] {\vspace*{0.6cm}\noindent{\normalsize\bf #1}
	\par\vspace*{0.4cm}}
					
\newcounter{appendixc}
\newcounter{subappendixc}[appendixc]
\newcounter{subsubappendixc}[subappendixc]

\renewcommand{\appendix}[1] {\vspace*{0.6cm}
        \refstepcounter{appendixc}
        \setcounter{figure}{0}
        \setcounter{table}{0}
        \setcounter{equation}{0}
        \renewcommand{\thefigure}{\Alph{appendixc}.\arabic{figure}}
        \renewcommand{\thetable}{\Alph{appendixc}.\arabic{table}}
        \renewcommand{\theappendixc}{\Alph{appendixc}}
        \renewcommand{\theequation}{\Alph{appendixc}.\arabic{equation}}
        \noindent{\bf Appendix \theappendixc #1}\par\vspace*{0.4cm}}

\def\abstracts#1{{
	\centering{\begin{minipage}{12.2truecm}\footnotesize\baselineskip=12pt\ 
noindent
	\centerline{\footnotesize ABSTRACT}\vspace*{0.3cm}
	\parindent=0pt #1
	\end{minipage}}\par}}

\renewenvironment{thebibliography}[1]
	{\begin{list}{\arabic{enumi}.}
	{\usecounter{enumi}\setlength{\parsep}{0pt}
\setlength{\leftmargin 1.25cm}{\rightmargin 0pt}
	 \setlength{\itemsep}{0pt} \settowidth
	{\labelwidth}{#1.}\sloppy}}{\end{list}}

\newcounter{itemlistc}
\newcounter{romanlistc}
\newcounter{alphlistc}
\newcounter{arabiclistc}

\newcommand{\fcaption}[1]{
        \refstepcounter{figure}
        \setbox\@tempboxa = \hbox{\footnotesize Fig.~\thefigure. #1}
        \ifdim \wd\@tempboxa > 6in
           {\begin{center}
        \parbox{6in}{\footnotesize\baselineskip=12pt Fig.~\thefigure. #1}
            \end{center}}
        \else
             {\begin{center}
             {\footnotesize Fig.~\thefigure. #1}
              \end{center}}
        \fi}

\newcommand{\tcaption}[1]{
        \refstepcounter{table}
        \setbox\@tempboxa = \hbox{\footnotesize Table~\thetable. #1}
        \ifdim \wd\@tempboxa > 6in
           {\begin{center}
        \parbox{6in}{\footnotesize\baselineskip=12pt Table~\thetable. #1}
            \end{center}}
        \else
             {\begin{center}
             {\footnotesize Table~\thetable. #1}
              \end{center}}
        \fi}

\def\@citex[#1]#2{\if@filesw\immediate\write\@auxout
	{\string\citation{#2}}\fi
\def\@citea{}\@cite{\@for\@citeb:=#2\do
	{\@citea\def\@citea{,}\@ifundefined
	{b@\@citeb}{{\bf ?}\@warning
	{Citation `\@citeb' on page \thepage \space undefined}}
	{\csname b@\@citeb\endcsname}}}{#1}}

\newif\if@cghi
\def\cite{\@cghitrue\@ifnextchar [{\@tempswatrue
	\@citex}{\@tempswafalse\@citex[]}}
\def\citelow{\@cghifalse\@ifnextchar [{\@tempswatrue
	\@citex}{\@tempswafalse\@citex[]}}
\def\@cite#1#2{{$\null^{#1}$\if@tempswa\typeout
	{IJCGA warning: optional citation argument
	ignored: `#2'} \fi}}

 1
 1
 1

\font\ninerm=cmr9

\begin{document}
\rightline{\footnotesize OUCMT-96-1}

\normalsize 
\thispagestyle{empty}
\setcounter{page}{1}


\vspace*{0.88truein}

\centerline{\bf MULTI-SPIN CODING OF THE MONTE CARLO SIMULATION}
\vspace*{0.035truein}
\centerline{\bf OF THE THREE-STATE RANDOM POTTS MODEL}
\vspace*{0.035truein}
\centerline{\bf AND THE BLOCK-SPIN TRANSFORMATION
\footnote{figures are available from
{\em http://glimmung.phys.sci.osaka-u.ac.jp/kikuchi/preprints.html}
or will be sent upon request.}}
\vspace*{0.37truein}
\centerline{\footnotesize MACOTO KIKUCHI\footnote
{E-mail: kikuchi@phys.sci.osaka-u.ac.jp}}
\vspace*{0.015truein}
\centerline{\footnotesize\it Department of Physics, Osaka University,
Toyonaka 560, Japan}
\vspace*{10pt}
\centerline{\normalsize and}
\vspace*{10pt}
\centerline{\footnotesize YUTAKA OKABE\footnote{E-mail:
okabe@phys.metro-u.ac.jp}}
\vspace*{0.015truein}
\centerline{\footnotesize\it Department of Physics, Tokyo Metropolitan
University, Tokyo 192-03, Japan}
\vspace*{0.225truein}

\vspace*{0.21truein}
\abstracts{
The multi-spin coding of the Monte Carlo simulation of the three-state
Potts model on the simple cubic lattice is presented.
The ferromagnetic (F) model, the antiferromagnetic (AF) model,
and the random mixture of the F and AF couplings are treated.
The multi-spin coding technique
is also applied to the block-spin transformation.
The block-spin transformation of the F Potts model is simply
realized by the majority rule, whereas the AF three-state Potts
model is transformed to the block spin having a six-fold symmetry.}{}{}

\vspace*{10pt}

\vspace*{1pt} 

\section{Introduction}
\noindent
Monte Carlo simulations are used as standard techniques to
investigate statistical mechanical properties of many-body systems.
In treating large systems, especially near the critical points
and at very low temperatures, we often encounter slow dynamics.
To overcome such slow dynamics,
the development of fast algorithms is demanded.
To gather more information from a single simulation
is another direction of effort for algorithmic improvement.
The cluster-flip Monte Carlo method,\cite{Swendsen87,Wolff89}
the histogram method,\cite{Ferrenberg88}  and the multicanonical
simulation\cite{Berg92} are examples of the recent progress.

For the simulations of the Ising model, where only one bit is required
for storing the information of a single spin, a computer word can store
several spins.  Based on this fact, the multi-spin coding technique has
been successfully developed.\cite{Creutz79,Zorn81,Williams84,Oliveira91}
Among them, the idea of the coding by Bhanot {\it et al.}\cite{Bhanot86}
and its variations are especially
useful.\cite{Michael86,Kikuchi87a,Kikuchi87b,Ito88}
In these methods, just a single bit is assigned to each spin.
Instead of storing spins at different sites in the same lattice in a
word, as in the case of former realizations of the multi spin coding,
Ising spins at the same site of several independent systems
(of the same lattice structure) are stored in a word.
The spins for 32 (64) systems are updated simultaneously in the case
of 32- (64-) bit machine, with a single random-number sequence;
for that to be achieved, all the operations are executed by logical
commands.  As a result, the computation time is reduced remarkably.
One can simulate either systems with different parameters, for example,
the temperature, the external field, etc., or systems under
the same parameters with different random number sequences.
Even for the latter purpose, one needs to generate only one
random-number sequence.\cite{Michael86,Kikuchi87a,Kikuchi87b,Ito93}
In the case of the simulations of random systems,
one can simulate systems with different
configurations in parallel.\cite{Kawashima93}
Since an average over a large number of random configurations
should be taken, the multi-spin coding technique is
particularly effective for random systems.
We should note that the multi-spin coding has also been applied to
the Monte Carlo simulations of the Ising model
on quasicrystals.\cite{Okabe88,Okabe90}

The Monte Carlo renormalization group (MCRG) is a powerful tool
to analyze critical phenomena.\cite{Swendsen79}  In MCRG method,
one performs the block-spin transformations at each Monte Carlo sweep.
The block-spin transformation makes a block spin from $b^D$ original spins,
with the scale factor $b$.  The present authors have
successfully used the multi-spin coding to the block-spin
transformation of the three-dimensional (3$D$) Ising
model.\cite{Kikuchi87a,Kikuchi87b}

The idea of the multi-spin coding can also be applied to other models
with discrete symmetries.  The $q$-state Potts model is one example.
The Potts model has various interesting properties.
depends
The order of the phase transitions of the ferromagnetic (F) Potts model depend
on the dimensionality, $D$, and the number of the state, $q$.
In 3$D$, the F Potts model shows a first-order transition, whereas
the antiferromagnetic (AF) model shows a second-order transition.
The random mixture of F and AF couplings gives rise to new problems:
The rounding of first-order transition in weak ferromagnetic region is
one of interesting problems;
one may also ask the universality of critical exponents in this case.
The Potts glass phase will be another problem of interest.
The zero-temperature transition of the three-state Potts glass has
been suggested in 3$D$.\cite{Scheucher90,Schreider94}

In this paper, we present the multi-spin coding of the three-state
Potts model on the simple cubic lattice.
For the coupling, we consider both cases of the F and AF couplings.
The random mixture of the F and AF couplings are also treated.
We also present the realization of the block-spin transformation
using multi-spin coding technique.  The appropriate choice of the
block-spin transformation is essential to extract the critical
properties of the order parameter.  We employ quite different
block-spin transformations for the F Potts model and the AF Potts model:
The block-spin transformation of the F model is simply realized by the
majority rule, whereas the block spin of the AF three-state Potts model
has the same symmetry as the six-clock model.

\section{Multi-Spin Coding}
\noindent
We are concerned with the three-state Potts Model, whose Hamiltonian
is given as
\begin{equation}
{\cal H} = - \sum_{<i,j>} J_{ij} \delta_{S_iS_j} \ ,
\end{equation}
where each spin $S_i$ can take three states, 0, 1 and 2.
Here a variable $J_{ij}$ may take either $+J$ or $-J$ generally.
In the following, we consider only the nearest-neighbor interactions.

We employ the multi-spin coding algorithm,\cite{Bhanot86,Michael86}
which was originally used for the Ising model.
In contrast to the Ising model,
two bits are required to represent a Potts spin;
the three states, 0, 1 and 2, are represented by (00), (01), and (10)
in binary-number representation.
Therefore, we assign one word for the upper bit and another word for
the lower bit, as is seen in the following.
It should be noted that the calculation of the order parameter becomes
quite simple in this two-word representation,
because the number of 1 (2) spin is counted just by summing up
the lower- (upper-) bit throughout the lattice.

To update a spin in one of the three states,
one has to choose either of the rest two states as a trial state.
This procedure is executed by the following code:
 
\begin{verbatim}
      iscoin=ira(la)
      isnew0=iand(not(isold0),iscoin)
      isnew1=not(ior(isold1,iscoin))
\end{verbatim}

\noindent
The old spin is represented by ({\tt isold1}, {\tt isold0}), with
the former word representing the upper bit and the latter the lower bit,
respectively.  Here, {\tt ira} is a random number,
where each bit takes 0 or 1 with the probability of 1/2.
Then, the trial state, also represented by two words,
({\tt isnew1}, {\tt isnew0}),
is selected as one of the other two states with an equal probability
for every case of  the old states (00),  (01), and (10) .

In the Metropolis algorithm,
one needs to count the local energy change due to the above trial
flip of a picked single spin.
The following code is for calculating the energy change (in unit of
$J$) at a bond:

\begin{verbatim}
      neqold = ior(ieor(isold0,is0),ieor(isold1,is1))
      neqnew = ior(ieor(isnew0,is0),ieor(isnew1,is1))
      ide0 = iand(neqold,neqnew)
      ide1 = not(neqnew)
\end{verbatim}

\noindent
Two words ({\tt is1},  {\tt is0}) represent the nearest-neighbor spin
in concern.  The energy change is also given by the binary-number
representation in two words, {\tt ide1} and {\tt ide0}.
The above code is for F Potts model.
For AF model, the last line should be modified as follows:

\begin{verbatim}
      ide1 = not(neqold)
\end{verbatim}

\noindent
And for the random-bond model, the same line should be read as

\begin{verbatim}
      ide1 = ior(iand(jbond,not(neqnew)),not(ior(jbond,neqold)))
\end{verbatim}

\noindent
where {\tt jbond} should either be $1$ for the ferromagnetic bond
or $0$ for the antiferromagnetic bond.
Since the coordination number is six for the simple cubic lattice, the local
energy change $W$ takes an integer value between $-6$ and 6 in units of $J$.
Thus, four words are necessary to represent $W$ in the multi-spin
coding.  It should be compared to the Ising case,
where three words are enough to represent the local energy change.

In using the random-number sequence,
we apply the trick proposed by Michael,\cite{Michael86}
rather than the original one by Bhanot {\it et al}.\cite{Bhanot86}
That is, we use the precalculated table for the transition
probability, $min[1, e^{-\beta J W}]$;
in the original algorithm, on the other hand,
the transition probability table is updated dynamically.
In case of using the precalculated table,
a large table is required to ensure the accuracy in temperature;
one can, however, use a very large table easily in computers these days,
so that it is not considered as a drawback.
On the other hand, using a small table which is dynamically updated
may cause an uncontrolable temperature fluctuation.

We use a dummy variable $X$, which takes the integer value from 0 to 6.
We prepare the table so that the probability
to get the value of $X$ is given by

  \[ X(r) = \left\{
           \begin{array}{lll}
              0,  & {\rm if} & r>e^{-K}           \\
              1,  & {\rm if} & e^{-K}>r>e^{-2K}   \\
              2,  & {\rm if} & e^{-2K}>r>e^{-3K}  \\
              3,  & {\rm if} & e^{-3K}>r>e^{-4K}  \\
              4,  & {\rm if} & e^{-4K}>r>e^{-5K}  \\
              5,  & {\rm if} & e^{-5K}>r>e^{-6K}  \\
              6,  & {\rm if} & e^{-6K}>r \ ,
           \end{array}
         \right. \]

\noindent
where $K=\beta J$ and $r$ is a random number which
uniformly takes the value between 0 and 1.
Then, the rule of the spin update will be as follows:
If $(-W)+X+8$ is greater than 8, then the spin will be flipped.

It should be emphasized that the shuffling of the precalculated
$X$-table is effective for simulating different samples with the same
parameters by a single random-number sequence.
The statistical dependence of random sequences in conjunction with
the shuffling of the table has been recently discussed.\cite{Ito93}

The FORTRAN code for the Metropolis method is given in Appendix,
which is for the F Potts model.  In our multi-spin coding,
$X$ is represented by three words, ({\tt jxtb2}, {\tt jxtb1},
{\tt jxtb0}).  We use the integer random number, {\tt irb},
which takes the value between 0 and {\tt nword-1}.  The decrease
in the local energy at the trial flip, $(-W)+8$, is calculated
to be ({\tt isum3}, {\tt isum2}, {\tt isum1}, {\tt isum0})
in binary-number representation.

The spin update is executed by the logical commands;
{\tt iflip} is the flag for the flip.
The updated spin and the resultant energy
difference are given by {\tt isp0} and {\tt iengd}, respectively.
With slight modifications, which are commented out with `{\tt *}',
we can also treat the AF model and the $\pm J$ model;
the information of the configuration of $\pm J$ bonds should be included
in {\tt jbond},
which in general depends on the lattice point {\tt la}.
For the convenience of vectorization, we divide the lattice into two
interpenetrating sublattices.

The linear size of the system is given by {\tt nx}; the number of
the spins on each sublattice is given by {\tt nla2=nx**3/2}.
The periodic boundary conditions are employed, and  informations on
the neighboring lattice points are provided in the tables,
{\tt jx}, {\tt jyr}, {\tt jyl}, {\tt jzr} and {\tt jzl}.

\section{Block-Spin Transformation}

\subsection{Ferromagnetic Order Parameter}
\noindent
For the three-state F Potts model,
it is convenient to express the order parameter as
a two-dimensional vector. Three states 0, 1, and 2 are expressed by the
vectors directing $\pm 2\pi /3$ from each other
as shown in the inset of Fig.~1.
\begin{figure}[htbp]
\vspace*{13pt}
\centerline{\vbox{\hrule width 5cm height0.001pt}}
\vspace*{6.3cm}
\centerline{\vbox{\hrule width 5cm height0.001pt}}
\vspace*{13pt}
\fcaption{Possible vectors obtained by the vector sum of eight F Potts spins.}
\end{figure}
%
Let us consider the case of the block-spin transformation from eight
spins; that is, the scale factor $b$ is 2.
Possible vectors obtained by the vector sum of
eight Potts spins are shown in Fig.~1.  From each vector,
we make a block spin which also takes
three states as does the original Potts spin;
we choose the Potts state directing closest
to the vector as the block-spin state.
The rule for determination of
the block spin is illustrated in Fig.~1,
where the separation is shown with solid lines.
At the border, the random number will be used to
determine which state is to be chosen.

One would find that the block-spin transformation described above
is, in fact, equivalent to the simple majority rule; in other words,
the three-state Potts spin is mapped onto the three-state Potts
spin by the majority rule transformation.

The essential part of the FORTRAN code for the block-spin transformation
is given as follows:

\begin{verbatim}
c
c       block-spin transformation
c           ferro 3-state Potts to 3-state Potts
c
      non02x = ior(isumx0,isumx1)
      nsumx2 = not(isumx2)
      non02y = ior(isumy0,isumy1)
      nsumy2 = not(isumy2)
      j0 = iand(iand(isumx0,isumx1),isumy1)
      j1 = iand(iand(isumy0,isumy1),isumx1)
c
      iran = ir(la)
      iranx = iand(isumx2,iran)
      irany = iand(isumy2,not(iran))
c
      is0 = iand(isumx2,iand(nsumy2,non02y))
      is0 = ior(is0,isumx3)
      is0 = ior(is0,iand(isumx2,non02x))
      is0 = ior(is0,iand(isumx2,irany))
      is0 = ior(is0,iand(nsumy2,iranx))
      ibnew0(la) = ior(is0,iand(j0,iran))
c
      is1 = iand(isumy2,iand(nsumx2,non02x))
      is1 = ior(is1,isumy3)
      is1 = ior(is1,iand(isumy2,non02y))
      is1 = ior(is1,iand(isumy2,iranx))
      is1 = ior(is1,iand(nsumx2,irany))
      ibnew1(la) = ior(is1,iand(j1,not(iran)))
\end{verbatim}

\noindent
Here, the sum of 0-bit of eight Potts spins is expressed by ({\tt isumx3},
{\tt isumx2}, {\tt isumx1}, {\tt isumx0}).
Similarly,
the sum of 1-bit is given by ({\tt isumy3}, {\tt isumy2},
{\tt isumy1}, {\tt isumy0}).
Two words, ({\tt ibnew1}, {\tt ibnew0}), gives the block spin.
The random number {\tt ir} is used for tie breaking at the border.
To make the MCRG analysis of the F order parameter,
we repeat the same block-spin transformation step by step.

\subsection{Antiferromagnetic Order Parameter}
\noindent
For the AF order parameter,
one should take account of the two-sublattice structure.
Three components of the order parameter are defined as follows:
\begin{equation}
q_{\sigma} \equiv (\sum_{i \in A} \delta_{S_i,\sigma} -
                  \sum_{j \in B} \delta_{S_j,\sigma})/N ,
\end{equation}
where $\sigma$ denotes one of the three states 0, 1 and 2, and
$A$ and $B$ stand for two sublattices.
Three states 0, 1, and 2 for the sublattices $A$ and $B$ can be expressed
by two-dimensional vectors as shown in the inset of Fig.~2.
\begin{figure}[htbp]
\vspace*{13pt}
\centerline{\vbox{\hrule width 5cm height0.001pt}}
\vspace*{6.6cm}
\centerline{\vbox{\hrule width 5cm height0.001pt}}
\vspace*{13pt}
\fcaption{Possible vectors obtained by the vector sum of eight AF Potts  
spins.}
\end{figure}
It should be noted here that each pair of the neighboring vectors
make an angle of $\pi/3$;
as a result, the order-parameter space in this case
has a six-fold symmetry
in contrast to that of the F Potts model.
This symmetry, $Z(6)$, is the same as that of the six-clock model.
Based on this observation,
we make block spins so as to preserve this six-fold symmetry;
more specifically, as the block spin we choose
one of the possible states of
the six-clock spin which directs closest to the vector sum
of the Potts spins in a block.
Possible vectors obtained by the vector sum of eight Potts spins
are illustrated in Fig.~2.
At the border, a block spin should be chosen in equal probability
by the random number.
At the center, on the other hand,
a block spin is chosen with the probability of 1/6.

We have chosen the primary
direction of the six-clock spin as shown in Fig.~2.
The reason is as follows:
We expect that the
ordered state of the AF three-state Potts model has
a broken-sublattice-symmetry order.\cite{Grest81,Wang90}
Namely, one sublattice is occupied by one of
the three states,
while the other sublattice is randomly occupied by the remaining two states.
The present choice of the spin direction takes this type of
order into account.

The block-spin transformation described above
for the AF order cannot be expressed as a simple majority rule
in contrast to the F case.
Therefore, we must perform the vector summation directly.
The multi-spin coding technique is applicable even in this case,
since all the operations can be executed only with integer variables
space.
when the oblique coordinates are used for expressing the order-parameter  
space.
The six states, 0, 1, 2, 3, 4, 5 in Fig.~2, are expressed
by (000), (001), (101), (100), (111), (011) respectively in our
binary-number representation.
We use the basis vectors ${\bf e}_x$ and ${\bf e}_y$ shown in Fig.~2.
Then, 0, 1 and 2 states on the sublattice $A$ are expressed
by ${\bf r}_A$, ${\bf e}_x + {\bf r}_A$ and ${\bf e}_y + {\bf r}_A$,
respectively,
where ${\bf r}_A = ({\bf e}_x + {\bf e}_y)/3$.
On the other hand, 0, 1 and 2 states on the sublattice $B$ are
expressed by ${\bf e}_x + {\bf e}_y + {\bf r}_B$,
${\bf e}_y + {\bf r}_B$ and ${\bf e}_x + {\bf r}_B$, respectively,
where ${\bf r}_B = 2({\bf e}_x + {\bf e}_y)/3$.

The FORTRAN code of the block-spin
transformation from the AF Potts spin to the six-clock spin
is given as follows:

\begin{verbatim}
c
c       block spin transformation
c           antiferro 3-state Potts to 6-clock
c
c  x>0 (icx = 0) or x=0 (icx = 1) or x<0 (icx = 2)
c   isumx >=< 4
c
      iwk = ior(isumx0,isumx1)
      icx0 = iand(isumx2,not(iwk))
      icx1 = not(ior(isumx2,isumx3))
c
c  y>0 (icy = 2) or y=0 (icy = 1) or y<0 (icy = 0)
c   isumy >=< 4
c
      iwk = ior(isumy0,isumy1)
      icy0 = iand(isumy2,not(iwk))
      icy1 = iand(ior(isumy2,isumy3),not(icy0))
c
c   isumy + isumx  (0-16)
c
      icr0 = iand(isumy0,isumx0)
      isumC0 = ieor(isumy0,isumx0)
      iwk = ieor(isumy1,isumx1)
      icr1 = ieor(iand(isumy1,isumx1),iand(icr0,iwk))
      isumC1 = ieor(icr0,iwk)
      iwk = ieor(isumy2,isumx2)
      icr2 = ieor(iand(isumy2,isumx2),iand(icr1,iwk))
      isumC2 = ieor(icr1,iwk)
      iwk = ieor(isumy3,isumx3)
      isumC3 = ieor(icr2,iwk)
      isumC4 = ieor(iand(isumy3,isumx3),iand(icr2,iwk))
c
c  y+x>0 (icm = 2) or y+x=0 (icm = 1) or y+x<0 (icm = 0)
c   isumy + isumx  >=< 8
c
      iwk = ior(isumC0,ior(isumC1,isumC2))
      icm0 = iand(isumC3,not(iwk))
      icm1 = iand(ior(isumC3,isumC4),not(icm0))
c
c  if border
c
      iran = ira(la)
c
      icb2 = ior(icm1,iand(icm0,iran))
      iwk1 = ior(icx1,iand(icx0,iran))
      iwk2 = ior(icy1,iand(icy0,iran))
      icb1 = iand(iwk1,iwk2)
      icb0 = not(ior(iwk1,iwk2))
c
c  if center
c
      icent = iand(icx0,iand(icy0,icm0))
      nocent = not(icent)
c
      ic2 = irb(la)
      ic1 = not(ior(irt1(la),iran))
      irt1(la) = ic1
      ic0 = iand(not(irt0(la)),iran)
      irt0(la) = ic0
c
      isblk2(la) = ior(iand(nocent,icb2),iand(icent,ic2))
      isblk1(la) = ior(iand(nocent,icb1),iand(icent,ic1))
      isblk0(la) = ior(iand(nocent,icb0),iand(icent,ic0))
\end{verbatim}

\noindent
The sum of the $x$- and $y$-components of eight Potts spins are given by
{\tt isumx} and {\tt isumy}.
The conditions needed for the determination of the block spin are
represented in terms of
the conditions on {\tt isumx}, {\tt isumy} and
{\tt isumx+isumy}, which are illustrated in Fig.~3.
\begin{figure}[htbp]
\vspace*{13pt}
\centerline{\vbox{\hrule width 5cm height0.001pt}}
\vspace*{3.4cm}
\centerline{\vbox{\hrule width 5cm height0.001pt}}
\vspace*{13pt}
\fcaption{The conditions needed for the determination of block spin are  
illustrated.}
\end{figure}
According to these conditions,
the three words to denote the six-clock spin,
({\tt isblk2}, {\tt isblk1}, {\tt isblk0}),
are determined by logical commands.
The choice of the state at the border is simply
made by the random number {\tt ira} with the probability of 1/2.
At the center, on the other hand,
we use two random numbers, {\tt ira} and {\tt irb}:
First, {\tt isblk2} is determined by the random number {\tt irb};
next, the same procedure is employed for
the determination of the three states represented by
{\tt isblk1} and {\tt isblk0}
as we used in choosing a trial state in the Monte Carlo update.
The information of the selected state,
({\tt irt1}, {\tt irt0}), is kept up to the next time.
With this procedure, we can pick up one state from six with
the equal probabilities.

Once the block spin having a six-fold symmetry is obtained,
the next step will be the transformation from the six-clock spin
to the block spin of the same symmetry.
The illustration of this transformation is given in Fig.~4.
\begin{figure}[htbp]
\vspace*{13pt}
\centerline{\vbox{\hrule width 5cm height0.001pt}}
\vspace*{6.5cm}
\centerline{\vbox{\hrule width 5cm height0.001pt}}
\vspace*{13pt}
\fcaption{Possible vectors obtained by the vector sum of eight six-clock  
spins.}
\end{figure}
%
Ties at the border and the center are treated similarly as before.
We can apply the multi-spin coding technique also to
the block-spin transformation from the six-clock spin to the six-clock block  
spin
as before.
Although we do not show the FORTRAN code for this transformation,
it is worthwhile to make one comment:
This time it is convenient
to use basis vectors ${\bf e}_{x'}$ and ${\bf e}_{y'}$ shown
in Fig.~4.
Then, the six states 0, 1, 2, 3, 4 and 5 are
represented respectively by $2{\bf e}_{x'} + {\bf e}_{y'} +
{\bf r}$, $2{\bf e}_{x'} + 2{\bf e}_{y'} + {\bf r}$,
${\bf e}_{x'} + 2{\bf e}_{y'} + {\bf r}$, ${\bf e}_{y'}
+ {\bf r}$, ${\bf r}$ and ${\bf e}_{x'} + {\bf r}$
with ${\bf r} = {\bf e}_{x'} + {\bf e}_{y'}$.
The conditions in terms of {\tt isumx-2*isumy}, {\tt 2*isumx-isumy}
and {\tt isumx+isumy} will be used for the determination of the block
six-clock state.

\section{Summary and Discussions}
\noindent
We have presented the multi-spin coding for the Monte Carlo simulation of
the three-state Potts model.
The multi-spin coding technique has also been applied
to the block-spin transformations of the Potts model both with the
F and AF order parameters.
The emphasis should be put on the fact that we have
totally excluded {\tt if}-statement in our code;
as a result, the code can perform ultrafast computations.

As we have mentioned in the Introduction,
some earlier applications of the multi-spin coding for the Ising model
\cite{Creutz79,Zorn81,Williams84} as well as a recent one\cite{Oliveira91}
showed how to store several spins belonging to the same lattice in a word.
This type of coding is effective for saving the memory of computers.
But not so dramatic acceleration in the simulation speed is expected.
On the other hand, the method we have presented
is based on yet another type of the multi-spin coding, which was
originally proposed by Bhanot {\it et al.}  for the Ising model\cite{Bhanot86}
and modified by Michael.\cite{Michael86}
In these methods and the present one as well, the multi-spin coding
technique is used in order that
a number of independent systems are simulated simultaneously.
These simultaneous simulation methods do not help reducing the memory  
consumption;
but we can expect a great reduction in the computation time.
Since the memory restriction is getting more and more relaxed these days,
the simultaneous simulation technique is advantageous over
the earlier multi-spin coding methods.
A simultaneous simulation method using 3 bits/spin
coding has also been used for the Ising model a decade ago.\cite{Kikuchi85}

A short comment will be made here on the performance of the present
code.  We have achieved the speed of 0.81 G Potts-spin flips per
second without any measurement in the case of 32-bits multi-spin
coding of the 3D Potts model for periodic boundary conditions
on the HITAC S-3800/480 system at the Computer Center of University
of Tokyo.  This is comparable to the speed of 2.10 G spin flips
per second for the 3D Ising model on the same machine:  For the 3D
Potts model the number of the logical commands for a spin update
is 110, which is about four times of that for the 3D Ising model.
With the measurement of the energy and the magnetization,
the performance for the 3D Potts model is changed to 0.49 G spin
flips per second:  When one makes block-spin transformations
three times with the measurement of block-spin magnetizations, the speed
becomes 0.44 G spin flips per second.  We should note that the present
code can be extended to the case of four-state Potts model, five-state
Potts model, and so on, although the coding will be more cumbersome.

Using this fast code, we have investigated several interesting topics of
the Potts model.
The order-parameter distribution functions of the F
three-state Potts model has been studied in 2D and 3D,\cite{Kikuchi92}
where attention has been paid to the vector character of the order
parameter.
We have also treated the 3D AF Potts model for a careful
study of the critical phenomena,
using the MCRG method combined with
finite-size scaling analysis;
we have confirmed that the 3D AF three-state Potts model is in the $XY$
universality class within statistical errors.\cite{Okabe92,Kikuchi95}

Another problem is the random three-state Potts model in 3D.
We have studied the $\pm J$ model with
general asymmetric probability weights.
The overall phase diagram of the random Potts model as a function of
the concentration of coupling and the temperature has been obtained.
In the AF-rich region,
we have paid attention to the universality of critical phenomena.
We have also shown the rounding of the first-order transition
due to the randomness in the F-rich region.\cite{Okabe93,Okabe95}

The details of the physical results are given in each
publication.\cite{Kikuchi92,Okabe92,Kikuchi95,Okabe93,Okabe95}

\nonumsection{Acknowledgments}
\noindent
This work is supported by a Grant-in-Aid for Scientific Research on
Priority Areas, ``Computational Physics as a New Frontier in Condensed
Matter Research'', from the Ministry of Education, Science and Culture,
Japan.

\nonumsection{References}

\appendix

\begin{verbatim}
c
c     Multi-Spin coding of the 3-dimensional 3-state Potts model
c           with ferromagnetic (F), antiferromagnetic (AF)
c                      and random couplings
c

      subroutine sbltce(isp0,isp1,jx,jyr,jyl,jzr,jzl,ira,irb,iengd)
      parameter(nx=64,nx2=nx/2,nxy2=nx2*nx,nla2=nxy2*nx)
      parameter(nword=2**24)
      common/jxtab/jxtb0(0:nword-1),jxtb1(0:nword-1)
     &            ,jxtb2(0:nword-1)
      dimension isp0(0:nla2-1,0:1),isp1(0:nla2-1,0:1)
      dimension jx(0:nla2-1),jyr(0:nla2-1),jyl(0:nla2-1)
     &                      ,jzr(0:nla2-1),jzl(0:nla2-1)
      dimension ira(0:nla2-1),irb(0:nla2-1)
      dimension iengd(0:nla2-1,0:3)

      do 10 la=0,nla2-1

c  old spin -> isold

        isold0 = isp0(la,0)
        isold1 = isp0(la,1)

c  set trial new spin -> isnew

        iscoin = ira(la)
        isnew0 = iand(not(isold0),iscoin)
        isnew1 = not(ior(isold1,iscoin))

c  energy difference for spin 1 -> ide1
c    neq : flag to tell whether n.n. spins are
c          the same (0) or not (1)
c    ide : e(old) - e(new) + 1
c      neqold < neqnew then ide = 0 (2) for F (AF)
c      neqold = neqnew then ide = 1
c      neqold > neqnew then ide = 2 (0) for F (AF)
c          Note that (neqold = 0) and (neqnew = 0) are not realized
c          at the same time.

        is0 = isp1(la,0)
        is1 = isp1(la,1)
        neqold = ior(ieor(isold0,is0),ieor(isold1,is1))
        neqnew = ior(ieor(isnew0,is0),ieor(isnew1,is1))
        ide10 = iand(neqold,neqnew)
*  for F :
        ide11 = not(neqnew)
*  for AF :
*        ide11 = not(neqold)
*  for random :
*        ide11 = ior(iand(jbond,not(neqnew)),not(ior(jbond,neqold)))

c  energy difference for spin 2 -> ide2

        is0 = isp1(jx(la),0)
        is1 = isp1(jx(la),1)
        neqold = ior(ieor(isold0,is0),ieor(isold1,is1))
        neqnew = ior(ieor(isnew0,is0),ieor(isnew1,is1))
        ide20 = iand(neqold,neqnew)
        ide21 = not(neqnew)
*        ide21 = not(neqold)
*        ide21 = ior(iand(jbond,not(neqnew)),not(ior(jbond,neqold)))

c  energy difference for spin 3 -> ide3

        is0 = isp1(jyr(la),0)
        is1 = isp1(jyr(la),1)
        neqold = ior(ieor(isold0,is0),ieor(isold1,is1))
        neqnew = ior(ieor(isnew0,is0),ieor(isnew1,is1))
        ide30 = iand(neqold,neqnew)
        ide31 = not(neqnew)
*        ide31 = not(neqold)
*        ide31 = ior(iand(jbond,not(neqnew)),not(ior(jbond,neqold)))

c  energy difference for spin 4 -> ide4

        is0 = isp1(jyl(la),0)
        is1 = isp1(jyl(la),1)
        neqold = ior(ieor(isold0,is0),ieor(isold1,is1))
        neqnew = ior(ieor(isnew0,is0),ieor(isnew1,is1))
        ide40 = iand(neqold,neqnew)
        ide41 = not(neqnew)
*        ide41 = not(neqold)
*        ide41 = ior(iand(jbond,not(neqnew)),not(ior(jbond,neqold)))

c  energy difference for spin 5 -> ide5

        is0 = isp1(jzr(la),0)
        is1 = isp1(jzr(la),1)
        neqold = ior(ieor(isold0,is0),ieor(isold1,is1))
        neqnew = ior(ieor(isnew0,is0),ieor(isnew1,is1))
        ide50 = iand(neqold,neqnew)
        ide51 = not(neqnew)
*        ide51 = not(neqold)
*        ide51 = ior(iand(jbond,not(neqnew)),not(ior(jbond,neqold)))

c  energy difference for spin 6 -> ide6

        is0 = isp1(jzl(la),0)
        is1 = isp1(jzl(la),1)
        neqold = ior(ieor(isold0,is0),ieor(isold1,is1))
        neqnew = ior(ieor(isnew0,is0),ieor(isnew1,is1))
        ide60 = iand(neqold,neqnew)
        ide61 = not(neqnew)
*        ide61 = not(neqold)
*        ide61 = ior(iand(jbond,not(neqnew)),not(ior(jbond,neqold)))

c  sum of energy differences for 2 spins

        isuma0 = ieor(ide10,ide20)
        isuma1 = ior(ieor(ide11,ide21),iand(ide10,ide20))
        isuma2 = iand(ide11,ide21)
        isumb0 = ieor(ide30,ide40)
        isumb1 = ior(ieor(ide31,ide41),iand(ide30,ide40))
        isumb2 = iand(ide31,ide41)
        isumc0 = ieor(ide50,ide60)
        isumc1 = ior(ieor(ide51,ide61),iand(ide50,ide60))
        isumc2 = iand(ide51,ide61)

c  isuma = isuma + 2

        isuma2 = ior(isuma1,isuma2)
        isuma1 = not(isuma1)

c  isumw = isuma + isumb

        isumw0 = ieor(isuma0,isumb0)
        icr = iand(isuma0,isumb0)
        iwk= ieor(isuma1,isumb1)
        isumw1 = ieor(iwk,icr)
        icr = ior(iand(isuma1,isumb1),iand(iwk,icr))
        iwk= ieor(isuma2,isumb2)
        isumw2 = ieor(iwk,icr)
        isumw3 = ior(iand(isuma2,isumb2),iand(iwk,icr))

c  isum = isumw + isumc
c    isum : total trial energy difference (old-new) +8
c       (2 <= isum <= 14)

        isum0 = ieor(isumw0,isumc0)
        icr = iand(isumw0,isumc0)
        iwk= ieor(isumw1,isumc1)
        isum1 = ieor(iwk,icr)
        icr = ior(iand(isumw1,isumc1),iand(iwk,icr))
        iwk= ieor(isumw2,isumc2)
        isum2 = ieor(iwk,icr)
        icr = ior(iand(isumw2,isumc2),iand(iwk,icr))
        isum3 = ior(isumw3,icr)

c  jxtb : X-table for Boltzmann factor
c       (0 <= jxtb <= 6)
c  iflip : flag for spin flip
c    isum + jxtb >= 8 then flip

        ir = irb(la)
        jt0 = jxtb0(ir)
        jt1 = jxtb1(ir)
        jt2 = jxtb2(ir)
        icr = iand(ieor(isum1,jt1),iand(isum0,jt0))
        icr = ior(iand(isum1,jt1),icr)
        icr = iand(ieor(isum2,jt2),icr)
        icr = ior(iand(isum2,jt2),icr)
        iflip = ior(isum3,icr)

c  new spin -> isp0(la)

        isp0(la,0) = ior(iand(iflip,isnew0),iand(not(iflip),isold0))
        isp0(la,1) = ior(iand(iflip,isnew1),iand(not(iflip),isold1))

c  energy difference (old-new) + 8  (2 <= iengd <= 14)

        iengd(la,0) = iand(iflip,isum0)
        iengd(la,1) = iand(iflip,isum1)
        iengd(la,2) = iand(iflip,isum2)
        iengd(la,3) = ior(not(iflip),isum3)

   10 continue

      end
\end{verbatim}


\begin{thebibliography}{99}

\bibitem{Swendsen87} R.~H.~Swendsen and J.~-S.~Wang,
{\it Phys. Rev. Lett.} {\bf 58}, 86 (1987).
\bibitem{Wolff89} U.~Wolff,
{\it Phys. Rev. Lett.} {\bf 62}, 361 (1989).
\bibitem{Ferrenberg88} A.~M.~Ferrenberg and R.~H.~Swendsen,
{\it Phys. Rev. Lett.} {\bf 61}, 2635 (1988).
\bibitem{Berg92} B.~A.~Berg and T.~Celik,
{\it Phys. Rev. Lett.} {\bf 69}, 2292 (1992).

\bibitem{Creutz79} M.~Creutz, L.~Jacobs and C.~Rebbi,
{\it Phys. Rev. Lett.} {\bf 42}, 1396 (1979).
\bibitem{Zorn81} R.~Zorn, H.~J.~Herrmann and C.~Rebbi,
{\it Comput. Phys. Commun.} {\bf 23}, 337 (1981).
\bibitem{Williams84} G.~O.~Williams and M.~H.~Kalos,
{\it J. Stat. Phys.} {\bf 37}, 283 (1984).
\bibitem{Oliveira91} P.~M.~C.~de~Olivieira,
{\it Computing Boolean Statistical Models} (World Scientific, 1991).
\bibitem{Bhanot86} G.~Bhanot, D.~Duke and R.~Salvador,
{\it Phys. Rev.} {\bf B33}, 7841 (1986);
{\it J. Stat. Phys.} {\bf 44}, 985 (1986).
\bibitem{Michael86} C.~Michael,
{\it Phys. Rev.} {\bf B33}, 7861 (1986).
\bibitem{Kikuchi87a} M.~Kikuchi and Y.~Okabe,
{\it Prog. Theor. Phys.} {\bf 78}, 540 (1987).
\bibitem{Kikuchi87b} M.~Kikuchi and Y.~Okabe,
{\it Phys. Rev.} {\bf B35}, 5382 (1987).
\bibitem{Ito88} N.~Ito and Y.~Kanada,
{\it Supercomputer} {\bf 25}, 31 (1988).
\bibitem{Kawashima93} N.~Kawashima, N.~Ito and Y.~Kanada,
{\it Int. J. Mod. Phys.} {\bf C4}, 525 (1993).
\bibitem{Okabe88} Y.~Okabe and K.~Niizeki,
{\it J. Phys. Soc. Jpn.} {\bf 57}, 16, 1536 (1988).
\bibitem{Okabe90} Y.~Okabe and K.~Niizeki,
{\it J. Phys.} {\bf A20}, L733 (1990).

\bibitem{Swendsen79} R.~H.~Swendsen,
{\it Phys. Rev. Lett.} {\bf 42}, 859 (1979).

\bibitem{Scheucher90} M.~Scheucher, J.~D.~Reger, K.~Binder
and A.~P.~Young,
{\it Phys. Rev.} {\bf B42}, 6881 (1990).

\bibitem{Schreider94} G.~Schreider and J.~D.~Reger,
{\it J. Phys.} {\bf A28}, 317 (1995).

\bibitem{Ito93} N.~Ito, M.~Kikuchi and Y.~Okabe,
{\it Int. J. Mod. Phys.} {\bf C4}, 569 (1993).

\bibitem{Grest81}
G.~S.~Grest and J.~R.~Banavar,
{\it Phys. Rev. Lett.} {\bf 46}, 1458 (1981).
\bibitem{Wang90} J.~-S.~Wang, R.~H.~Swendsen and R.~Koteck\'y,
{\it Phys. Rev.} {\bf B42}, 2465 (1990).

\bibitem{Kikuchi85} M.~Kikuchi and Y.~Okabe,
{\it Prog. Theor. Phys.}{\bf 74}, 458 (1985).
\bibitem{Kikuchi92} M.~Kikuchi and Y.~Okabe,
{\it J. Phys. Soc. Jpn.} {\bf 61}, 3503 (1992).
\bibitem{Okabe92} Y.~Okabe and M.~Kikuchi,
in {\it Computational Approaches in Condensed-Matter Physics},
eds. S.~Miyashita, M.~Imada and H.~Takayama,
(Springer, Berlin, 1992), p. 193.
\bibitem{Kikuchi95} M.~Kikuchi and Y.~Okabe,
in preparation.
\bibitem{Okabe93} Y.~Okabe and M.~Kikuchi,
in {\it Computer Aided Innovation of New Materials II},
eds. M.~Doyama, J.~Kihara, M.~Tanaka and R.~Yamamoto,
(Elsevier, Amsterdam, 1993), p. 429.
\bibitem{Okabe95} Y.~Okabe and M.~Kikuchi,
in {\it Computational Physics as a New Frontier in Condensed Matter Research},
eds. H.~Takayama, M.~Tsukada, H.~Shiba, F.~Yonezawa, M.~Imada and Y.~Okabe,
(The Physical Society of Japan, Tokyo, 1995), p. 247.

\end{thebibliography}
\end{document}